\documentclass[jkps,preprint,fleqn,showpacs,showkeys,10pt]{revtex4}
\usepackage{graphicx}
\usepackage{amssymb}
\usepackage{amsmath}
\usepackage{bm}

\usepackage{latexsym,amsfonts}
\usepackage[dvipsnames]{xcolor}
\usepackage{epsfig}
\usepackage[all]{xy}
\begin{document}

\setcounter{page}{0}
\title[]{How Can We Erase States Inside a Black Hole?}
\author{Junha \surname{Hwang}}
\email{hwangjh@postech.ac.kr}
\affiliation{Department of Physics, POSTECH, Pohang 37673, Republic of Korea}
\author{Hyosub \surname{Park}}
\email{adad561@dgist.ac.kr}
\affiliation{School of Undergraduate Studies, College of Transdisciplinary Studies,  
Daegu Gyeongbuk Institute of Science and Technology (DGIST), Daegu 42988, Republic of Korea}
\author{Dong-han \surname{Yeom}}
\email{innocent.yeom@gmail.com}
\affiliation{Asia Pacific Center for Theoretical Physics, Pohang 37673, Republic of Korea}
\affiliation{Department of Physics, POSTECH, Pohang 37673, Republic of Korea}
\author{Heeseung \surname{Zoe}}
\email{heezoe@dgist.ac.kr}
\affiliation{School of Undergraduate Studies, College of Transdisciplinary Studies,  
Daegu Gyeongbuk Institute of Science and Technology (DGIST), Daegu 42988, Republic of Korea}


\begin{abstract}
We investigate an entangled system, which is analogous to a composite system of a black hole and Hawking radiation. If Hawking radiation is well approximated by an outgoing particle generated from pair creation around the black hole, such a pair creation increases the total number of states. There should be a unitary mechanism to reduce the number of states inside the horizon for black hole evaporation. Because the infalling antiparticle has negative energy, as long as the infalling antiparticle finds its partner such that the two particles form a separable state, one can trace out such a zero energy system by maintaining unitarity. In this paper, based on some toy model calculations, we show that such a unitary tracing-out process is only possible before the Page time while it is impossible after the Page time. Hence, after the Page time, if we assume that the process is unitary and the Hawking pair forms a separable state, the internal number of states will monotonically increase, which is supported by the Almheiri-Marolf-Polchinski-Sully (AMPS) argument. In addition, the Hawking particles cannot generate randomness of the entire system; hence, the entanglement entropy cannot reach its maximum. Based on these results, we modify the correct form of the Page curve for the remnant picture. The most important conclusion is this: if we assume unitarity, semi-classical quantum field theory, and general relativity, then the black hole should violate the Bekenstein-Hawking entropy bound around the Page time at the latest; hence, the infinite production arguments for remnants might be applied for semi-classical black holes, which seems very problematic.
\end{abstract}



\maketitle

\section{Introduction}

The information loss problem of black holes \cite{Hawking:1976ra} is one of the unresolved problems of quantum gravity, and the tension between general relativity and quantum mechanics needs to be reconciled. This problem is related to diverse topics of physics, including gravity, semi-classical quantum field theory, various approaches of quantum gravity, thermodynamics and statistical mechanics, and information theory. Among these topics, sometimes a toy model based on information theory gives important insights into the information paradox.

One of the most important examples from the information theoretical approach was investigated by Page \cite{Page:1993df}. Following this pioneering work, one considers a bipartite system where one part is an analogy of a black hole and the other part is an analogy of Hawking radiation \cite{Hawking:1974sw}. Initially, one assumes that all particles are in the black hole. As time goes on, one takes particles from one side and moves them to the other side. By assuming pure and random configurations, one can estimate the behavior of the Boltzmann entropy and the entanglement entropy. Eventually, one observes that after almost half of the initial Boltzmann entropy is emitted, the outside observer can distinguish the Boltzmann entropy from the entanglement entropy; i.e., one can notice that the quantum state of radiation is no more exactly thermal. Therefore, after the halfway point, or the so-called Page time, one can distinguish information from radiation.

If the Boltzmann entropy is proportional to the black hole area, i.e., if the Boltzmann entropy is the same as the Bekenstein-Hawking entropy \cite{Bekenstein:1973ur}, then the Page argument shows that information should be emitted after the halfway point, where the area of the black hole can be still large. Hence, a reasonable guess is that Hawking radiation should carry information \cite{Susskind:1993if}. However, once Hawking radiation carries information, various inconsistencies are already known to appear \cite{Yeom:2008qw,Almheiri:2012rt}. One way to overcome the troubles is to suppose that the Boltzmann entropy is not proportional to the black hole area \cite{Chen:2014jwq}. Then, we open a possibility that a small-sized black hole can carry many degrees of freedom. Such a possibility can be well described by introducing tripartite systems. The authors have investigated the Page curves\footnote{The Page curve means a Boltzmann entropy versus entanglement entropy relation for each part when there are more than two subsystems.} for tripartite systems \cite{Hwang:2016otg}, where the first part is the black hole, the second part is radiation, and the third part is the so-called (broadly defined) remnant. The authors newly observed that information can be carried in the form of mutual information, though it is determined by the number of states of the remnant.

For both approaches (bipartite and tripartite systems), what we have considered is to move particles from here to there. However, one may ask the question: \textit{does the Hawking radiation anything other than move a particle from inside to outside the horizon}? The answer is not so simple. A more correct interpretation is to say that Hawking radiation is a kind of pair creation, where one particle comes out and the other particle comes in; hence, first, the outgoing particle is entangled with its ingoing counterpart, and second, the ingoing counterpart interacts with the inside (collapsed matter).

This is not the end of the story. If this happens, then the total number of states is purely increased because two particles have been added. However, what one usually wants is to erase a number of states inside the horizon. Then, \textit{can one do that in a unitary way}? In terms of energy, one may think that this is fine because the infalling antiparticle will carry negative energy; i.e., the infalling counterpart corresponds to the negative energy flux toward the black hole \cite{Davies:1976ei}. However, if two particles are erased (say, a negative energy particle and a positive energy particle) from the system; i.e., if two particles are traced out from the system, unitarity will be lost and the state will evolve to a mixed state.

The only possibility to trace out two particles in a unitary way is to consider an un-entangled pair; i.e., the antiparticle (negative energy) and the particle (positive energy, which is identified as Hawking radiation) form a \textit{separable state}. Then, can an infalling antiparticle of Hawking radiation always find \textit{its partner inside the black hole} to form a separable state again? Of course, the opposite process is usual; i.e., an evolution from a separable state to an entangled state is natural. However, if the interaction is rapid enough, then a separable pair may possibly be returned from the entangled system as long as it is, in principle, possible. The problem is, whether this is possible in principle or not.

The purpose of this paper is to find a toy model that reveals the limitations of the information emission process via entangled particle pairs. In order to do this, we investigate several toy models based on spin-$1/2$ particles. We can divide the system and can calculate the entanglement entropy as usual. Also, we can bring another system and turn on the interaction between two parts; hence, we can increase the entanglements. One important assumption behind this is that the entanglement entropy or quantum states of black holes should be well approximated by such a random state or random interactions.

What we conclude is that such a unitary tracing-out procedure inside the black hole is allowed before the Page time, but disallowed after the Page time. Based on numerical calculations, we explicitly justify the result, and this point is our new contribution to the context.

Then, as long as Hawking particle-antiparticle pairs are separable systems, the number of states of inside and outside the horizon should increase as evaporation goes on because no unitary mechanism to erase the internal states of the black hole exists. In addition to this, the interaction channel is not sufficient to randomize inside and outside the horizon after the Page time. Therefore, our conclusion is that this system is no longer a random system so the Page formula will not be guaranteed. Finally, we will compare our results with various opinions in the literature and find applications in the context of the information loss problem.

This paper is organized as follows: In Section II, we describe toy models for interacting bipartite systems. Based on these models, in Section III, we discuss our hypothetical scenarios that realize unitary evaporation. Finally, in Section IV, we critically comment on possible obstacles toward the ultimate understanding of the information loss problem.

\section{Randomization of interacting systems}\label{sec:toy}

In this section, we investigate several toy models of interacting bipartite systems. First, we introduce two systems $A$ and $B$, where $A$ has $n$ states and $B$ has $m$ states. According to the Page formula, if we assume a pure and random system, the entanglement entropy between $A$ and $B$, i.e., $S(A|B)$, is approximately $\log n$ if $n < m$ and $\log m$ if $m < n$.

In the original work of Page \cite{Page:1993df}, one simply moves particles from $A$ to $B$ in order to mimic evaporation. However, in reality, the situation is intermediated by Hawking radiation. In order to investigate various cooperations of Hawking particles, we consider several numerical experiments. After the descriptions, we comment on physical meanings and implications.

\subsection{Preliminaries: quantum states and entanglements}

In order to use toy models, we need the following three kinds of operations. First, we define a density matrix for a random state. Second, for a given system, we add one more separable system and mix these systems by using the swapping operation, which will be explicitly defined later. Third, from the mixed combination, by tracing out the density matrix, we calculate the entanglement entropy. In this subsection, we discuss how to define a random state, how to entangle two systems, how to randomize two systems, and how to trace out the density matrix, where these methods are the same as those in \cite{Hwang:2016otg}.

We consider three subsystems of spin-$1/2$ particles, which we call $A$, $B$, and $C$, respectively. The number of particles for each part are $N$, $M$, and $L$, respectively. The number of states for each part are $n=2^{N}$, $m=2^{M}$, and $\ell=2^{L}$, respectively, where $X = n \times m \times \ell$ is the total number of states. Then, an arbitrary quantum state can be represented by
\begin{eqnarray}
| \psi_{A \cup B \cup C} \rangle = \sum_{i_{1},..., i_{N+M+L}} c_{i_{1},...,i_{N+M+L}} |i_{1},...,i_{N+M+L}\rangle,
\end{eqnarray}
where $i_{j} = 1,2$ (up or down), $|i_{1},...,i_{N+M+L}\rangle$ are orthonormal basis, and $c_{i_{1},...,i_{N+M+L}}$ are complex numbers that satisfy the normalization condition $\langle \psi | \psi \rangle = 1$. If the values of $c_{i_{1},...,i_{N+M+L}}$ are assigned by random numbers, then this state is \textit{random}.

Let us consider a situation in which $A$ has a state given by
\begin{eqnarray}
| \psi_{A} \rangle = \sum_{i_{1},..., i_{N}} c_{i_{1},...,i_{N}} |i_{1},...,i_{N}\rangle
\end{eqnarray}
with random numbers $c_{i_{1},...,i_{N}}$, and $B$ has a state given by
\begin{eqnarray}
| \psi_{B} \rangle = \sum_{i_{N+1},..., i_{N+M}} c_{i_{N+1},...,i_{N+M}} |i_{N+1},...,i_{N+M}\rangle
\end{eqnarray}
with random numbers $c_{i_{N+1},...,i_{N+M}}$. Then, $A$ and $B$ are random states. Now, we can consider a composite system that consists of $A$ and $B$:
\begin{eqnarray}
| \psi_{A\cup B} \rangle = \sum_{i_{1},..., i_{N+M}} c_{i_{1},...,i_{N+M}} |i_{1},...,i_{N+M}\rangle,
\end{eqnarray}
where
\begin{eqnarray}
|i_{1},...,i_{N+M}\rangle = |i_{1},...,i_{N}\rangle \otimes |i_{N+1},...,i_{N+M}\rangle
\end{eqnarray}
and
\begin{eqnarray}
c_{i_{1},...,i_{N+M}} =  \alpha \left( i_{1},...,i_{N+M} \right) c_{i_{1},...,i_{N}} c_{i_{N+1},...,i_{N+M}},
\end{eqnarray}
with $\alpha \left( i_{1},...,i_{N+M} \right)$ being a complex number. If $\alpha \left( i_{1},...,i_{N+M} \right) = 1$ for all $i$'s, then we can present the total state as a direct product of $|\psi_{A}\rangle$ and $|\psi_{B}\rangle$; hence, $A$ and $B$ are \textit{separable}. If the total wave function cannot be presented as a direct product of two wave functions, then this system is not separable, and the two subsystems are \textit{entangled}. Using this technique, we can consider a quantum state for the union of $A$, $B$, and $C$.

One may turn on interactions between subsystems. In this paper, we use a series of random swap operations. For example, for a given state $|\psi_{A\cup B\cup C} \rangle$, let us define a swap operation between $A$ and $B$ only. We define a unitary (orthogonal) matrix $\mathcal{O}$ that acts $|\psi_{A\cup B\cup C} \rangle$ by
\begin{eqnarray}
\mathcal{O}^{(i^*,j^*)} = \mathrm{SW}_{A \cup B}^{i^*:j^*} \otimes \mathcal{I}_{C}
\end{eqnarray}
for $i^* \in [i_1, i_N], j^* \in [i_{N+1},i_{N+M}]$, 
where $\mathrm{SW}_{A \cup B}$ is an $(nm \times nm)$-matrix whose elements are 
\begin{eqnarray}
&& \mathrm{SW}_{A \cup B}^{i^*:j^*}[i,j] = \delta_{ij} \;\;\;\;\;\;\;\; \mathrm{for}\;\; i,j \neq i^*,j^*,\\
&&\mathrm{SW}_{A \cup B}^{i^*:j^*}[i^*,i^*] = \mathrm{SW}_{A \cup B}^{i^*:j^*}[j^*,j^*] = 0,\\
&& \mathrm{SW}_{A \cup B}^{i^*:j^*}[i^*,j^*] = \mathrm{SW}_{A \cup B}^{i^*:j^*}[j^*,i^*] = 1,
\end{eqnarray}
and $\mathcal{I}_{C}$ is an $(\ell \times \ell)$-identity matrix. 
$\mathrm{SW}_{A \cup B}^{i^*:j^*}$ looks complicated but is just a typical swap gate in quantum computations. If we choose $i^*$ and $j^*$ randomly and repeat this process many times, then we realize a series of \textit{random swap operations}.

In order to measure the entanglements, we first define the density matrix $\rho$:
\begin{eqnarray}
\rho_{A \cup B \cup C} = | \psi_{A \cup B \cup C} \rangle \langle \psi_{A \cup B \cup C} |.
\end{eqnarray}
One can trace out a part of the total system. For example, by tracing out the degrees of freedom of $A$, we obtain the density matrix for $B \cup C$:
\begin{eqnarray}
\rho_{B \cup C} = {\mathrm{tr}}_{A} \rho_{A\cup B\cup C},
\end{eqnarray}
where in terms of components,
\begin{eqnarray}
\rho_{B \cup C}[i,j] = \sum_{k=1}^{n} \rho_{A\cup B\cup C}[m\ell(k-1)+i,m\ell(k-1)+j].
\end{eqnarray}
With the same method, we can define any combination of density matrices of subsystems. Finally, we can define the entanglement entropy between $B \cup C$ and $A$:
\begin{eqnarray}
S(B\cup C|A) = - \mathrm{tr} \rho_{B\cup C} \log \rho_{B\cup C}.
\end{eqnarray}
If $A\cup C$ and $B$ are separable, then the entanglement entropy is zero; if entangled, then it has a non-zero value.

\begin{figure}
\begin{center}
\includegraphics*[scale=0.5,viewport=50 210 500 600]{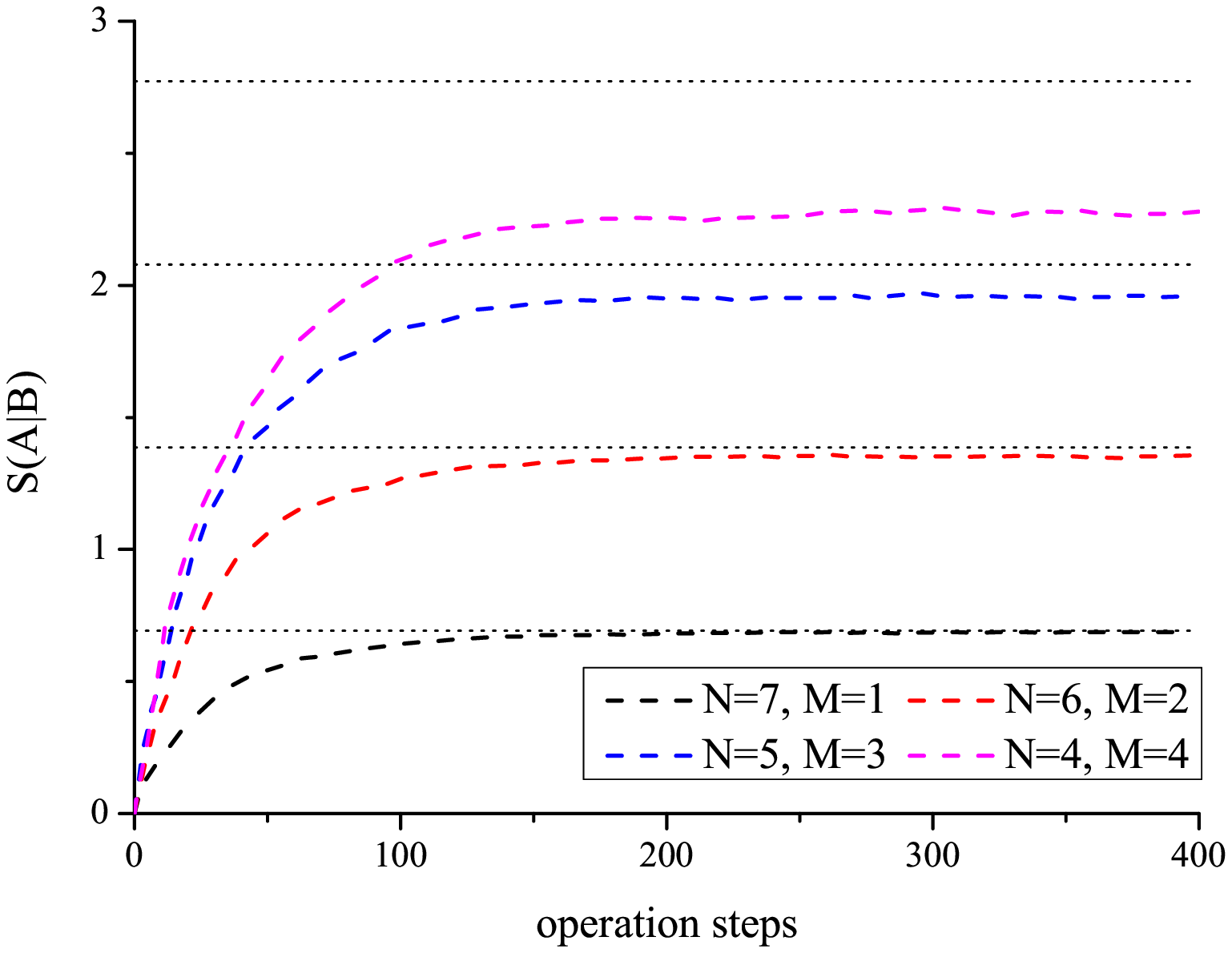}
\includegraphics*[scale=0.5,viewport=50 210 500 600]{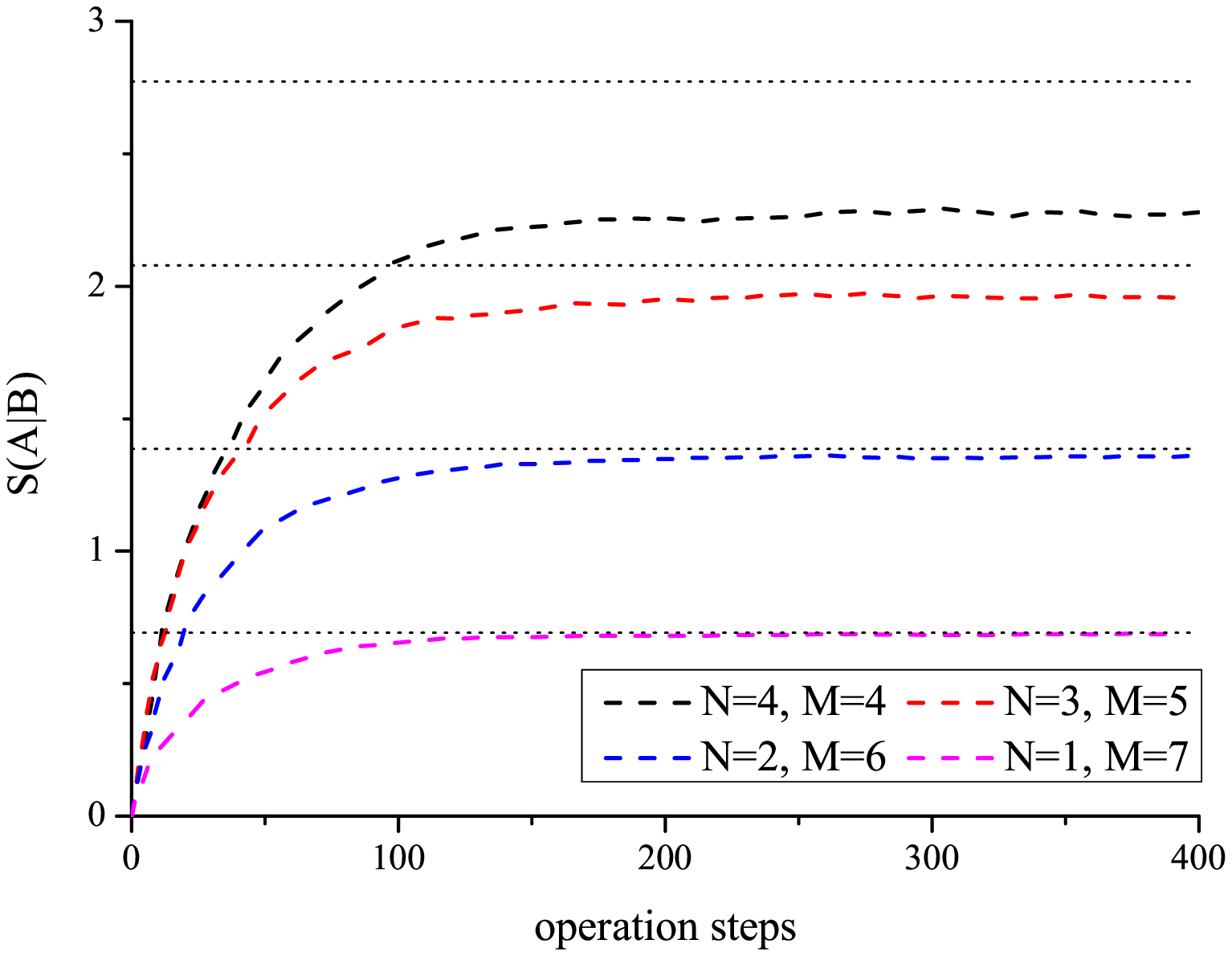}
\caption{\label{fig:sim1}$S(A|B)$ as the operations are accumulated. Left is for before the Page time ($N \geq M$), and right is for after the Page time ($N \leq M$). Horizontal lines are $\log 2$, $2 \log 2$, $3 \log 2$, and $4 \log 2$ from bottom to top.}
\end{center}
\end{figure}

\subsection{Toy model experiments}

We describe numerical experiments for our toy models. Because numerical data can be different at each trial due to random swap operations, we have averaged the results $10$ times for every experiment.

\subsubsection{Model 1: via random interactions}

$A$ is defined to have $N$ particles and $B$ to have $M$ particles, where $N + M$ is a fixed number. For spin-$1/2$ particles, the total number of states for $A$ is $n = 2^{N}$, and that for $B$ is $m = 2^{M}$. We start from the initial quantum state which is random for $A$ and $B$, respectively, but no entanglement exists between $A$ and $B$; i.e., $A$ and $B$ are initially separable. From this initial condition, we define the swapping operation between $A$ and $B$. Now our purpose is to check whether this operation can randomize the total system or not.

As a measure of the randomness, we calculate the entanglement entropy between $A$ and $B$, i.e., $S(A|B)$, by varying $N$ and $M$. According to the Page limit, if $N < M$, the entanglement entropy should be $\simeq N \log 2$; if $M < N$, then it should be $\simeq M \log 2$. Numerically, we can check that the entanglement approaches the Page limit (Figure~\ref{fig:sim1}) if $N$ and $M$ have a hierarchy ($N \ll M$ or $N \gg M$), which is consistent with the conjecture.

\subsubsection{Model 2: via partial interactions}

Now let us assume that $A$ and $B$ are randomized and entangled with each other. From this initial condition, we add two particles that are separable in the beginning, say subsystem $C$. We now turn off the interactions between $A$ and $B$ and turn on the interactions between $A$ and $C$ only. Our purpose is to calculate the entanglement entropy $S(A \cup C|B)$ and to check whether this one-sided interaction can randomize the system and maintain the Page formula or not. If it is not randomized, then there may be a bias from the Page formula.

From the beginning, because $C$ is separable, $S(A \cup C|B) = S(A|B)$. As we turn on the interactions, if we still follow the Page formula, then the entanglement entropy will not be changed for $N > M$ while the entanglement entropy should increase up to $(N+2) \log 2$ for $N < M$ (more correctly, for $N+2 < M$).

\begin{figure}
\begin{center}
\includegraphics*[scale=0.5,viewport=50 210 500 600]{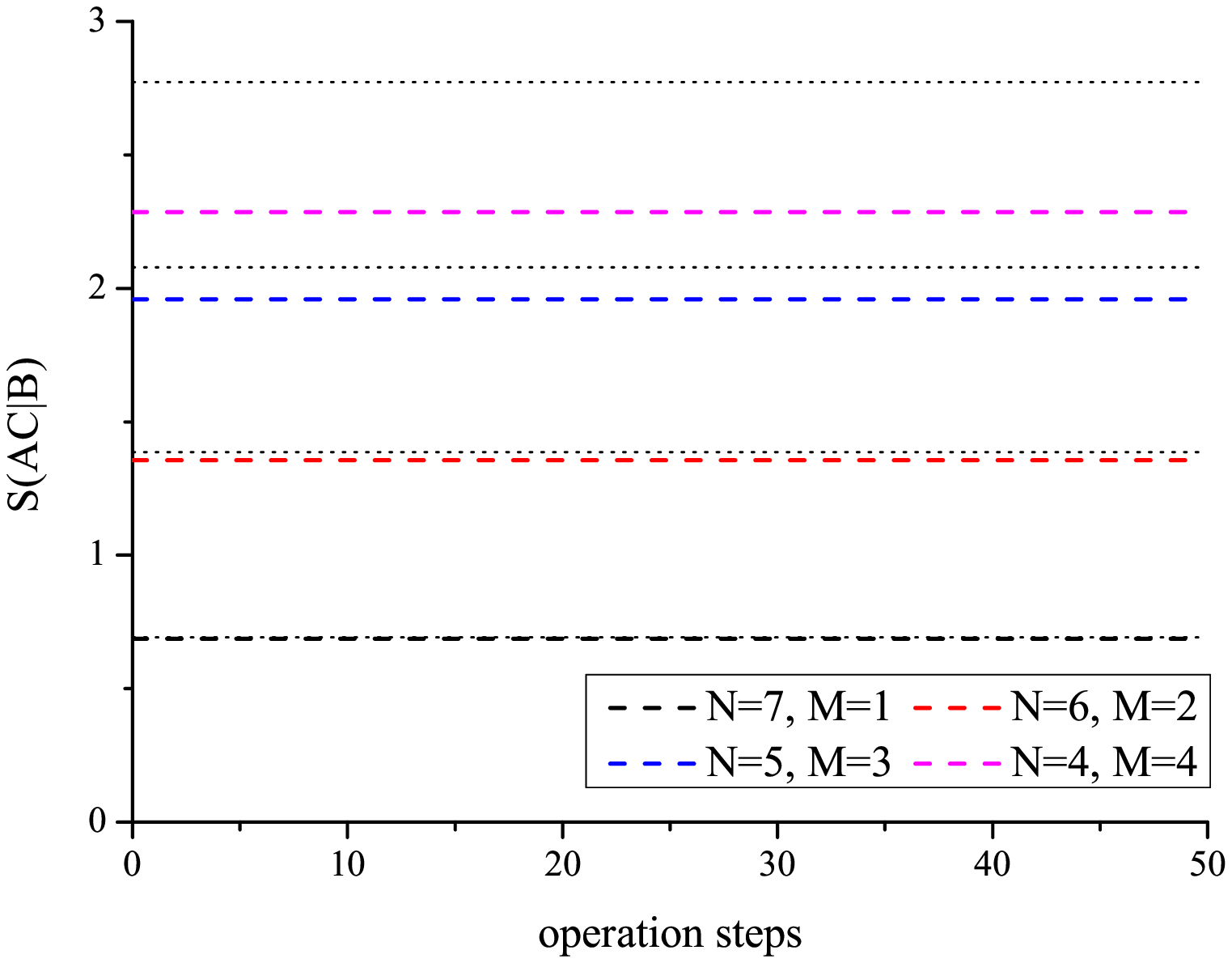}
\includegraphics*[scale=0.5,viewport=50 210 500 600]{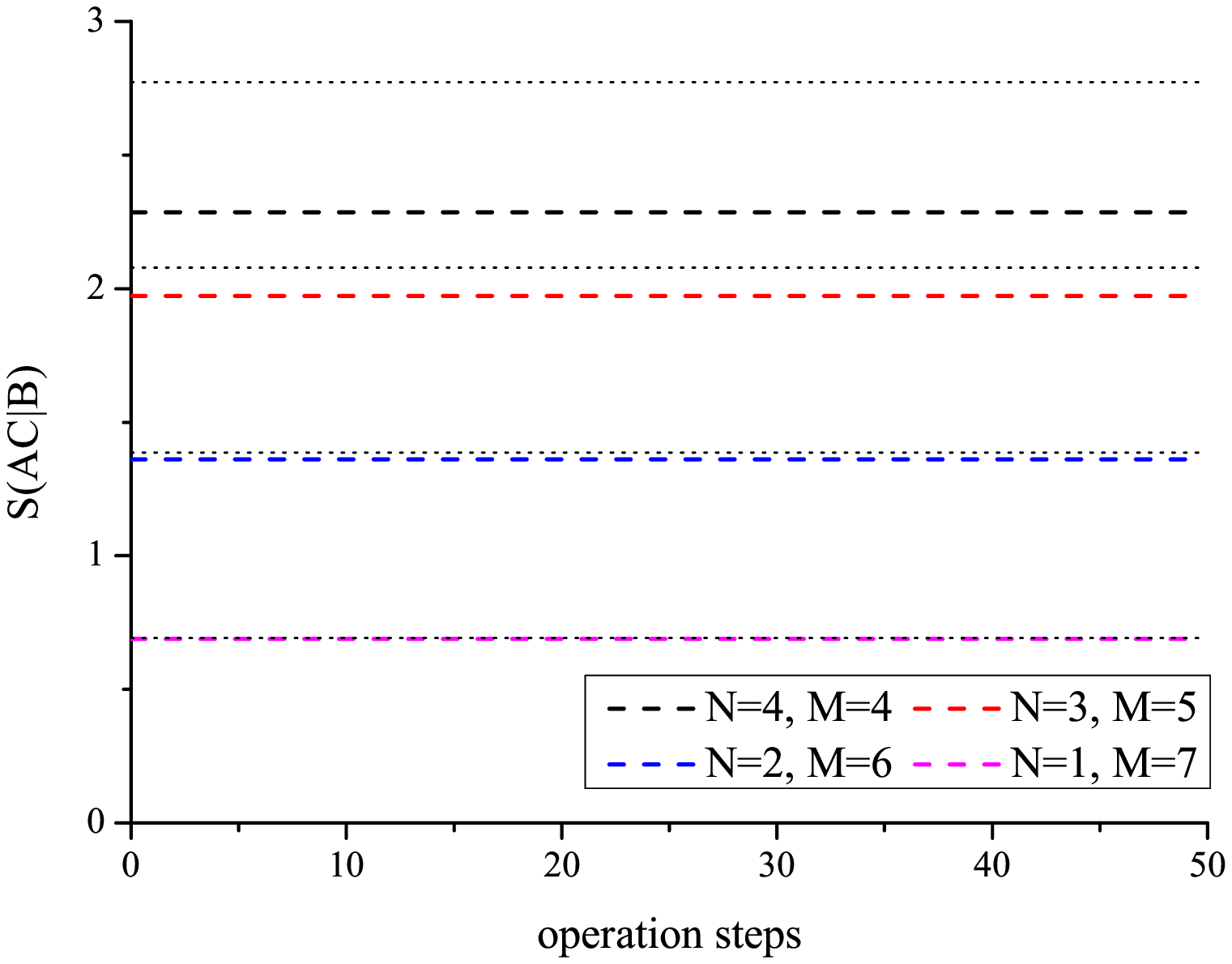}
\caption{\label{fig:sim2}$S(A\cup C|B)$ as the operations are accumulated. Left is for before the Page time ($N \geq M$), and right is for after the Page time ($N \leq M$).}
\end{center}
\end{figure}

Our numerical observations confirm that the entanglement entropy does not change if two particles interact with only one side (Figure~\ref{fig:sim2}). This means that the Page formula needs to be biased if $N < M$, where $C$ interacts with only $N$ particles of $A$. This may be interpreted as follows: if $N > M$, even though two particles are only interacting with $A$, no serious bias from the entire randomness exists because the number of states for $A$ is much larger than that for $B$. On the other hand, if $N < M$, the interaction with only $A$ cannot randomize the entire system.

This result has three important conclusions. Let us assume that $A$ is a black hole and $B$ is radiation outside the black hole. Then $m < n$ corresponds to a time before the Page time and $n < m$ corresponds to a time after the Page time. Keeping this in mind, we pressed the following conclusions:
\begin{itemize}
\item[C1.] Even though $A$ and $B$ are randomized, if $n < m$ and if one adds a separable system $C$ and only turns on the interaction with $A$, then the total system cannot maintain the randomness. After the Page time, one may see that the entanglement entropy between the black hole and the radiation no longer follows the Page formula.
\item[C2.] Before the Page time, one can (approximately) randomize $C$ by an interaction with $A$. On the other hand, after the Page time, one cannot randomize $C$ by an interaction with $A$. We summarize this as follows: before the Page time, a unitary operation $U$ exists such that
\begin{displaymath}
\xymatrix{\mathrm{separable}\;\; A \cup C \;\; \ar[r]^{U\;\;\;\;\;\;\;\;\;\;\;\;\;\;\;} & \;\; \mathrm{configuration\;with\;the\;Page\;limit},
}
\end{displaymath}
while this is impossible after the Page time.
\item[C3.] Because all operations are unitary, one may say oppositely that \textit{one can find a separable $C$ from $A$ before the Page time while one cannot find such a separable subsystem after the Page time.} We again summarize this as follows: before the Page time, a unitary operation $U^{-1}$ exists such that
\begin{displaymath}
\xymatrix{\mathrm{configuration\;with\;the\;Page\;limit} \;\; \ar[r]^{\;\;\;\;\;\;\;\;\;\;\;\;\;\;\;U^{-1}} & \;\; \mathrm{separable}\;\; A \cup C,
}
\end{displaymath}
while this is impossible after the Page time.
\end{itemize}
This opposite process is, of course, probabilistically disfavored while randomization is the preferred direction. However, if one can allow enough time for operations (on the order of the Poincare recurrence time), then such a recurrence will be possible. We will discuss the physical meaning of finding a separable $C$ later, but what we want to emphasize is that such a finding is impossible (even if we wait for the Poincare recurrence) if the system is after the Page time.

\begin{figure}
\begin{center}
\includegraphics*[scale=0.5,viewport=50 210 500 600]{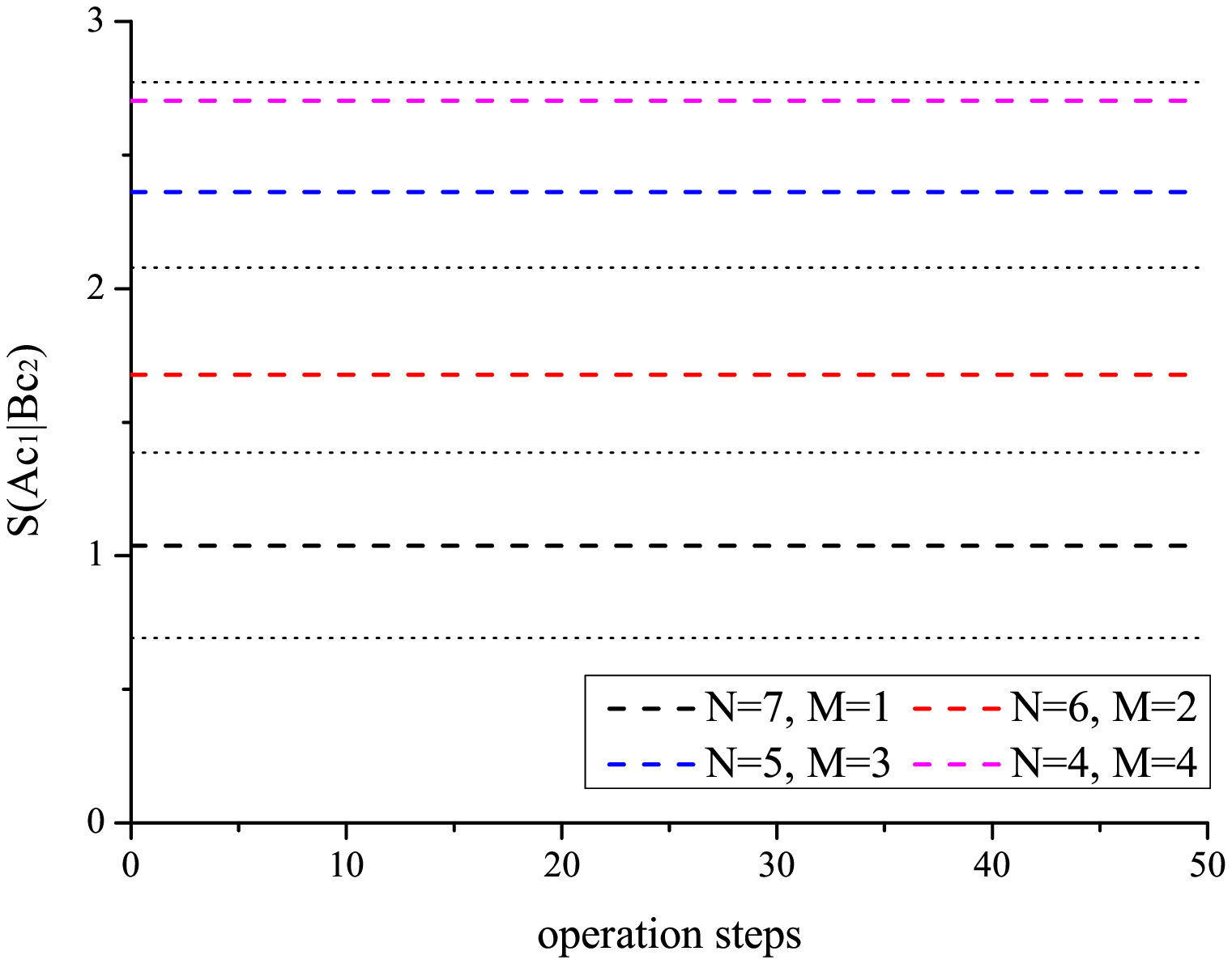}
\includegraphics*[scale=0.5,viewport=50 210 500 600]{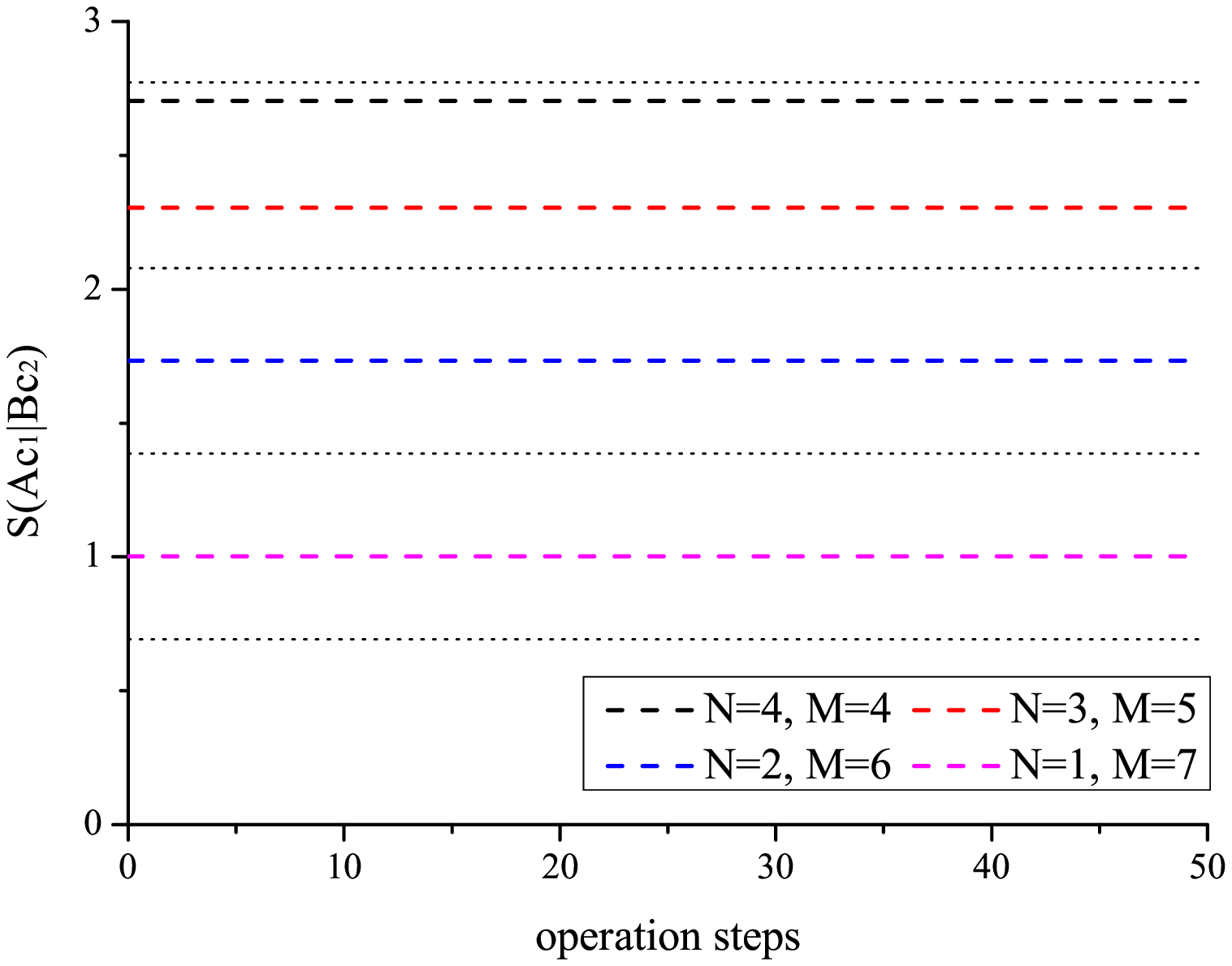}
\caption{\label{fig:sim3}$S(A\cup c_{1}|B\cup c_{2})$ for case 1 as the operations are accumulated. Left is for before the Page time ($N \geq M$), and right is for after the Page time ($N \leq M$). Compared to Figure~\ref{fig:sim1}, entanglement entropies are increased. However, they do not vary as time goes on and never approach the Page limit even for $n \ll m$ or $n \gg m$.}
\includegraphics*[scale=0.5,viewport=50 210 500 600]{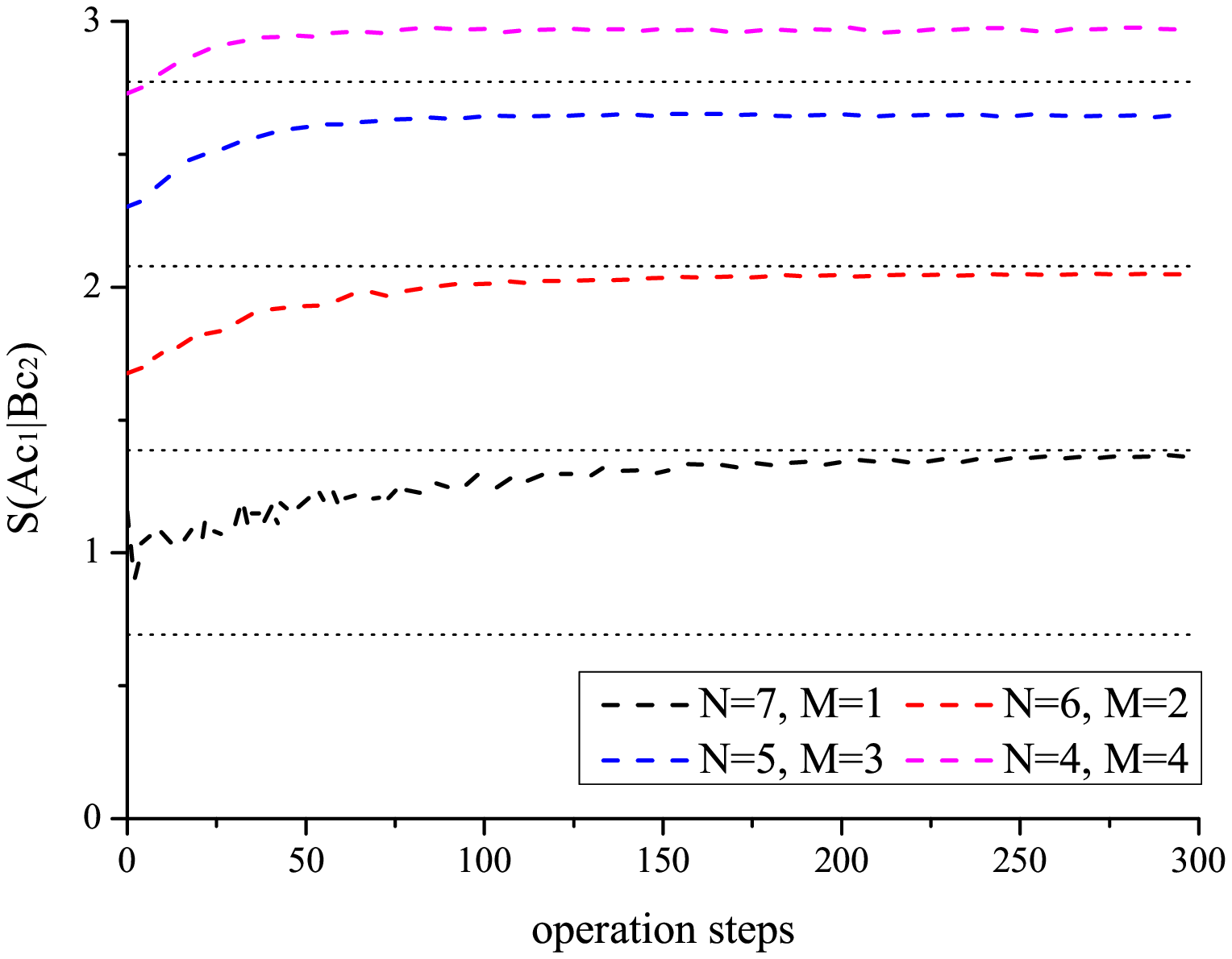}
\includegraphics*[scale=0.5,viewport=50 210 500 600]{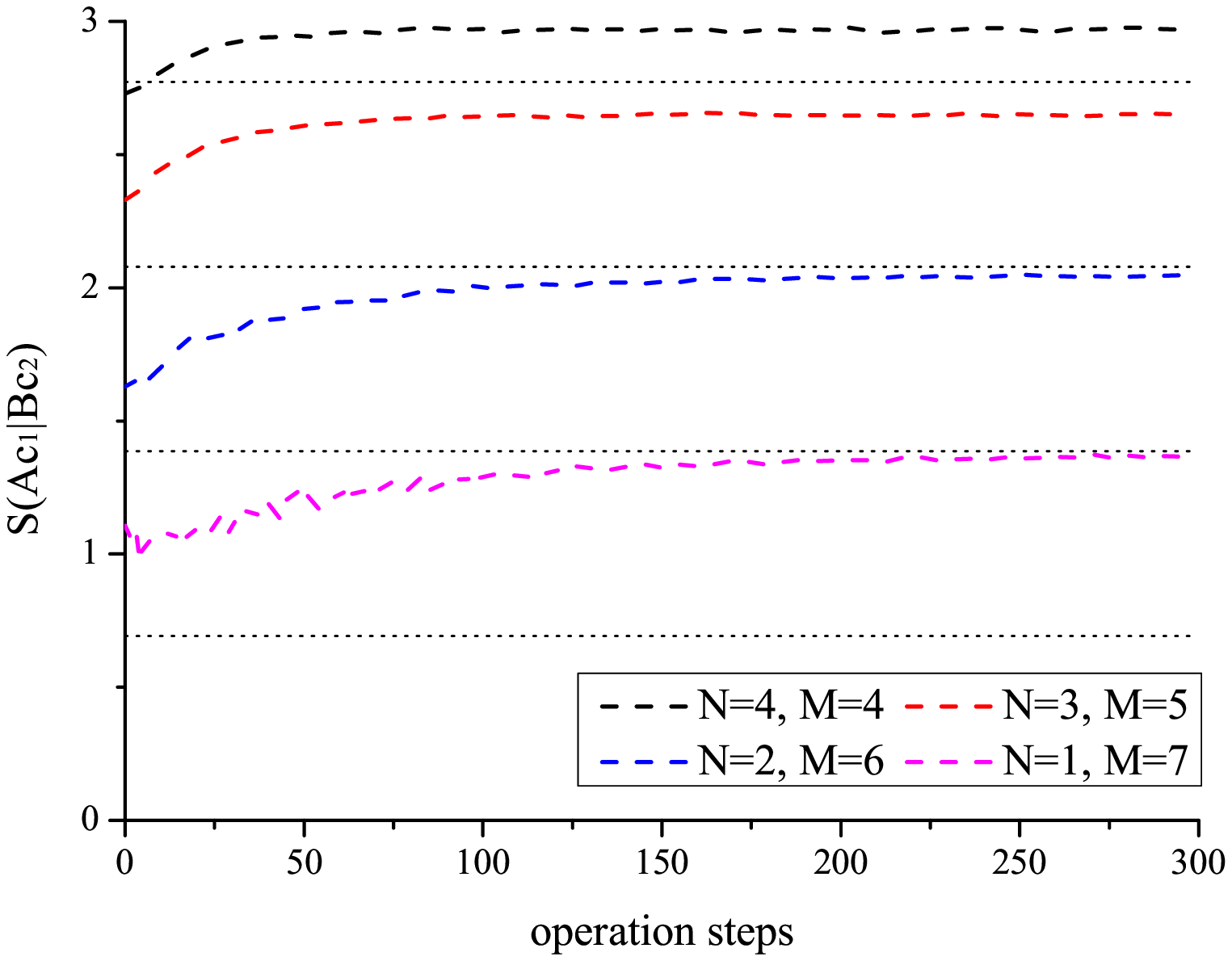}
\caption{\label{fig:sim32}$S(A\cup c_{1}|B\cup c_{2})$ for case 2 as the operations are accumulated. Left is for before the Page time ($N \geq M$), and right is for after the Page time ($N \leq M$). Due to this non-local interactions, the entanglement entropy approaches the Page limit.}
\end{center}
\end{figure}

\subsubsection{Model 3: via intermediators}

Let us consider $A$ and $B$ again. The total system is pure and random. From this, we introduce $C$, which is composed of two particles and is initially separable from $A$ and $B$. Now, we turn on the interaction $C$. We consider the following two cases:
\begin{itemize}
\item[--] Case 1. One particle of $C$ (say, $c_{1}$) is interacting with $A$ while the other particle (say, $c_{2}$) is not.
\item[--] Case 2. Both $c_{1}$ and $c_{2}$ are randomly interacting with $A$ and $B$ while no direct interaction taken place between $A$ and $B$.
\end{itemize}
Now, we calculate the entanglement entropy $S(A\cup c_{1}|B\cup c_{2})$ for both cases.

We may find an analogy such that $A$ is a black hole, $B$ is radiation, and $C$ is a particle($c_{2}$)-antiparticle($c_{1}$) pair of Hawking radiation. In usual cases, $c_{1}$ only interacts with $A$ (Case 1). However, one may further guess that the new particle-antiparticle pair will allow non-local interactions between $A$ and $B$ (Case 2).

Numerical observations are interesting. If we only consider local interactions (Case 1), then after one particle falls in, the entanglement entropy in Fig.~\ref{fig:sim3} is slightly increased compared to Fig.~\ref{fig:sim1}. However, it does not approach the correct Page formula, and the entropy never changes even though interactions are continued inside the black hole. This means that the addition of separable Hawking pairs will increase the entanglement entropy, but it cannot maintain the randomness of the entire system.

However, if we turn on non-local interactions, then the entanglement entropy can further increase and approach the Page limit (Fig.~\ref{fig:sim32}). Therefore, if non-local interactions exist, then the Page limit of entanglements can be maintained and perhaps so can the randomness of the entire system.

\section{A scenario for unitary evaporation}\label{sec:sec}

Following the pioneering work of Page, we may consider the black hole and radiation as a bipartite system. As time goes on, states go from inside to outside. If all processes are unitary, then the entanglement entropy follows the Page curve, and after the halfway point, one can distinguish information from radiation.

However, in realistic situations, this is not so simple. The true situation is something like the follows. First, Hawking radiation is generated as the result of a particle-antiparticle pair creation. The particles in this pair are entangled with each other, but the relation to the internal degrees of freedom is less clear. As the antiparticle goes into the black hole and carries negative energy, it should \textit{erase} one particle inside the black hole; this is the only way to maintain the total number of degrees of freedom. Then, how can this be possible by unitary quantum mechanics?

\begin{figure}
\begin{center}
\includegraphics[scale=0.45]{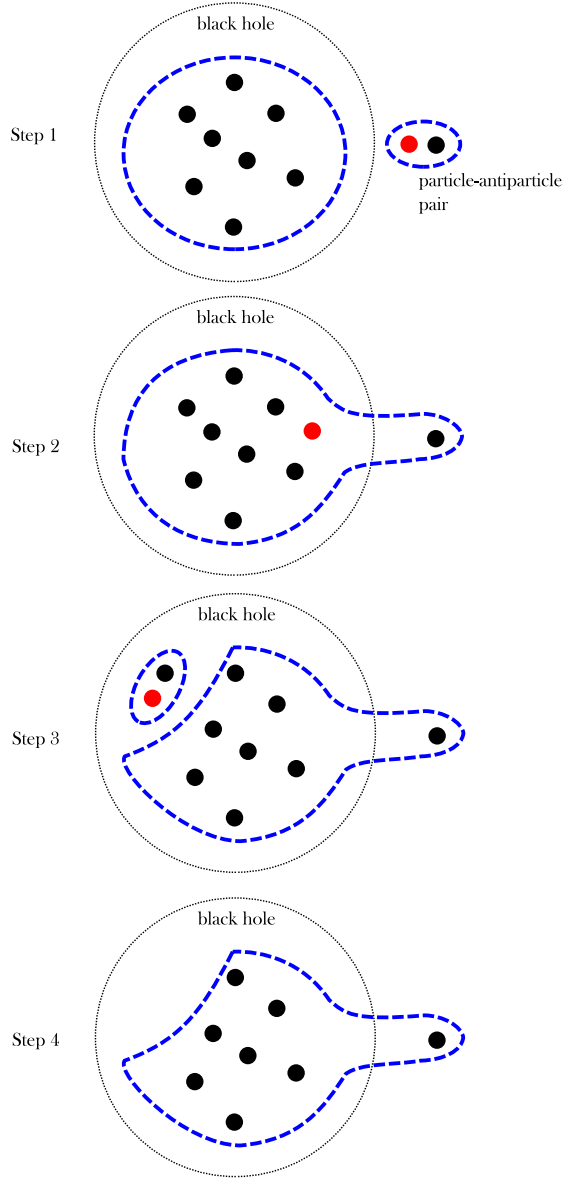}
\caption{\label{fig:Fig}A conceptual picture of unitary evaporation. Black dotted circles are the boundary of the black hole, and blue dashed curves are the boundary of entanglements. Step 1: a particle-antiparticle pair is generated. Step 2: the antiparticle (with negative energy) falls into the horizon and starts to interact. Step 3: after some interactions, the antiparticle can find a partner such that the antiparticle-partner pair forms a separable state. Step 4: Then, we can unitarily trace out the two separable particles, where the pair has zero energy. In the end, our interpretation is that one particle comes from the black hole in a unitary way.}
\end{center}
\end{figure}

\subsection{Before the Page time: can we remove a number of states from inside the black hole?}

In this subsection, we introduce a scenario in which a number of states are removed from inside the black hole, as summarized in Fig.~\ref{fig:Fig}. For all figures, the black dotted circles are the boundary of the black hole, and the blue dashed curves are the boundary of entanglements. If two blue dashed curves are disconnected, which means that the two systems are separable and have no entanglements.

In the first step, a black hole exists with a given number of states. In addition to this, we create a particle (black) and antiparticle (red) pair. Then, the total number of degrees of freedom apparently increases because the number of particles increases. We make a further simplification that the black dot has positive energy and the red dot has negative energy\footnote{Since the time coordinate inside the horizon is spacelike, one may interpret this as negative momentum; however, we can still say that this is a negative energy particle based on the fact that there is an incoming negative energy flux due to the renormalized energy-momentum tensors.} where the total energy summation of the particle-antiparticle pair is zero in order to satisfy energy conservation. In this figure, the blue dashed curves (boundary of entanglements) are disconnected because we have no reason to think that the created particles are entangled with the internal degrees of freedom. 

In the second step, the antiparticle falls into the black hole and starts to interact with the interior of the black hole. Then (as we demonstrated), the entanglement entropy of the particle-antiparticle pair will increase as the number of operations or interactions increases. 

In the third step, the antiparticle continuously interacts with the other particles. As time goes on, the antiparticle may possibly find a partner such that the antiparticle-partner combination forms a separable state\footnote{Of course, the particle and the antiparticle will be tightly entangled. Here, we emphasize that the particle and antiparticle pair is separated from the other systems, e.g., the black hole.} (hence, the blue dashed curves are disconnected) as we have shown in the previous section. This is not natural, but in principle possible; if a separable state evolves into an entangled state, then its inverse is always possible, though the price is that such a probability is very low. Hence, in realistic situations, a very long time of interactions is required; $\mathcal{T} \simeq \delta t \times \mathcal{N}$, where $\delta t$ is the unit time of each interaction and $\mathcal{N}$ is the necessary number of interactions for such a separation.

In the last step, we suppose that tracing-out the separated antiparticle-partner pair is censored. This pair has two properties: (1) quantum states are separable from the others and (2) the total energy is zero. Therefore, this means that we can erase two particles without violating unitarity or energy conservation. Then, in the end, the total number of degrees of freedom of this black hole and radiation system is preserved while one particle comes out from the black hole and this outgoing Hawking particle maintains the entanglements with the internal degrees of freedom. If this process is repeated, then it is equivalent to moving particles from inside to outside. 

For each step had several simplifications. For example, we make the simplification that Hawking radiation is nothing but a particle-antiparticle pair creation. Also, can we assume that such an antiparticle with a negative energy has an independent degree of freedom? These are still unresolved, but on the other hand, they do not look to be so unreasonable. However, apart from this, the too most important assumptions are as follows:
\begin{itemize}
\item[--] 1. \textit{Rapid mixing}: As the antiparticle falls into the black hole, rapid interactions take place between the collapsed matter and the antiparticle, satisfying the condition that $\mathcal{T}$ would be reasonably small.
\item[--] 2. \textit{Unitarity censorship}: The antiparticle can and must be traced out only if it finds its partner that satisfies a separable combination.
\end{itemize}
By accepting these two assumptions, we can see that this scenario will work.

We understand that these processes require these two strong assumptions and need justifications. However, what we can say is that this searching for a separable antiparticle-partner system is the \textit{only} way to trace out and erase degrees of freedom inside the horizon. If the previous assumptions are satisfied, then this mechanism can be maintained up to the Page time; if they are not, then the number of states will purely increase even before the Page time.

\begin{figure}
\begin{center}
\includegraphics[scale=0.75]{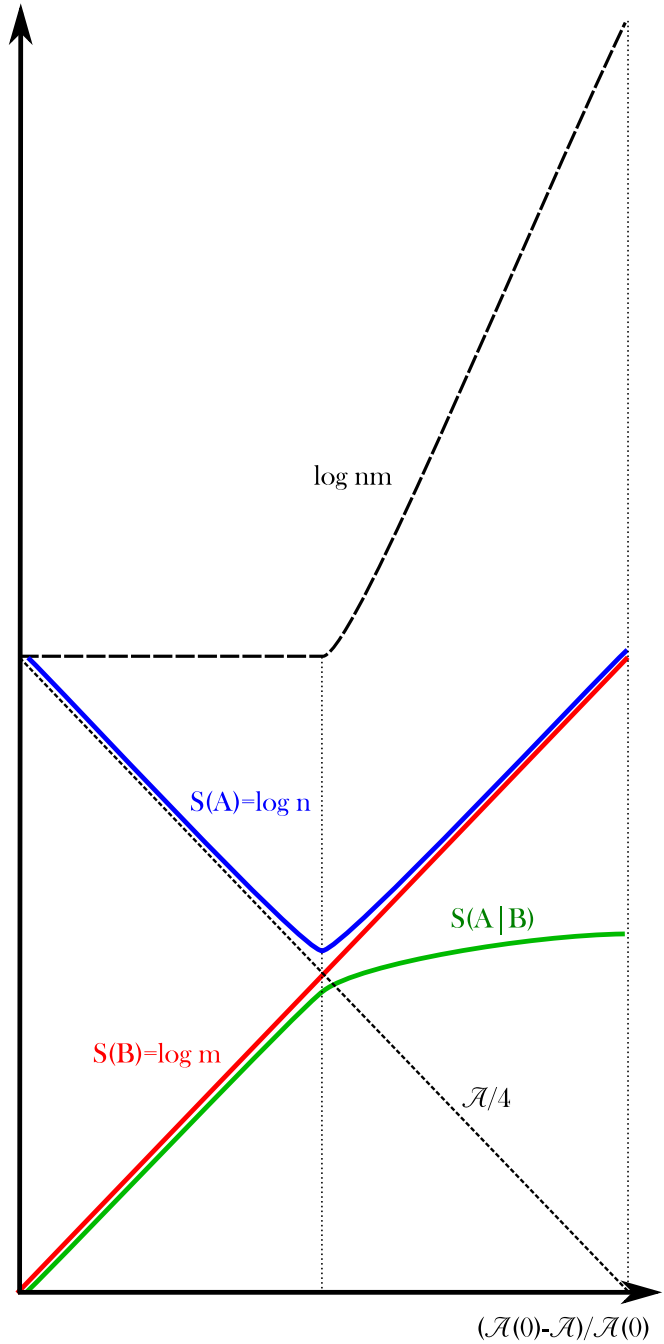}
\caption{\label{fig:newpage}The modified Page curve based on unitary evolutions, where we assume that before the Page time (thin dotted line), the total number of states (thick black dashed curve) is conserved. After the Page time, erasing the number of states inside the black hole is impossible; hence, the total number of states should increase and eventually become twice the initial value. The state after the Page time cannot be random; hence, $S(A|B)$ (green) is much lower than $S(A)$ (blue) or $S(B)$ (red). Assuming Hawking's formula, the areal entropy $\mathcal{A}/4$ (black dotted) will monotonically decrease. Hence, at the latest, after the Page time, the Bekenstein-Hawking entropy bound should break down.}
\end{center}
\end{figure}

\subsection{After the Page time: can the Page limit be maintained?}

The discussion of the previous subsection will work only if one can find such an antiparticle-partner pair. However, what happens if the antiparticle cannot find its partner? As we have seen in the previous section, even though the time is long enough, if the number of internal degrees of freedom is much less than the number of total degrees of freedom, then the mixing of the internal degrees of freedom can be restricted. Therefore, such a partner particle cannot be found.

We can see this from another point of view. Let us consider a situation in which a black hole has reached the Page time, say $N_{1} = M_{1}$ at $t_{1}$. If one adds a separable system and turns on interactions with only one side, then the entanglement entropy between the black hole and the radiation never changes (as we see in Fig.~\ref{fig:sim2}), say $S(A|B) \simeq N_{1} \log 2$. However, if the infalling antiparticle can find its partner and can be annihilated unitarily at time $t_{2} > t_{1}$, then the degrees of freedom of the black hole will be decreased, and we can see $N_{2} < N_{1} = M_{1} < M_{2}$ (as Step 4 of Fig.~\ref{fig:Fig}). If this is the case, at $t_{2}$, the entanglement entropy between the black hole and the radiation should be decreased because $S(A|B) \leq S(A) = N_{2}\log 2 < N_{1} \log 2$. However, we previously mentioned that the entanglement entropy of radiation cannot be changed because we included just a separable system with only one side. Therefore, this is inconsistent; hence, after the Page time, the antiparticle cannot be annihilated.

\begin{figure}
\begin{center}
\includegraphics[scale=0.45]{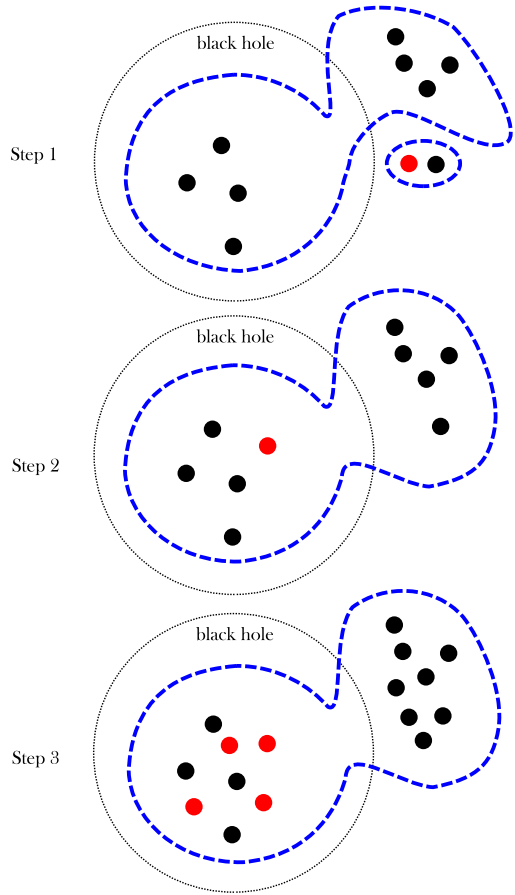}
\caption{\label{fig:Fig_mod}After the Page time, an antiparticle may not be able to find its partner. Step 1: a particle-antiparticle pair is generated. Step 2: the antiparticle falls into the horizon and cannot find its proper partner. Step 3: in the end, the total degrees of freedom is increased by a factor of two, where radiation is outside and a combination of antiparticles with half the original degrees of freedom is inside.}
\end{center}
\end{figure}

Note that this is directly related to the AMPS (Almheiri-Marolf-Polchinski-Sully) thought experiment \cite{Almheiri:2012rt}. According to this thought experiment, three subsystems are introduced, where the first one is the earlier part of Hawking radiation (say, $\mathcal{E}$) that was emitted before the Page time, the second one is the later part of Hawking radiation (say, $\mathcal{L}$), and the third one is the infalling counterpart (corresponding antiparticles) of $\mathcal{L}$ (say, $\mathcal{F}$).

Then, the following inequalities should be satisfied:
\begin{eqnarray}
&&S_{\mathcal{EL}} + S_{\mathcal{LF}} \geq S_{\mathcal{L}} + S_{\mathcal{ELF}},\\
&&|S_{\mathcal{E}} - S_{\mathcal{LF}}| \leq S_{\mathcal{ELF}} \leq S_{\mathcal{E}} + S_{\mathcal{LF}}.
\end{eqnarray}
As we assumed, if Hawking particles are separable from the beginning, then one can choose $S_{\mathcal{LF}} = 0$. In addition, after the Page time, the entanglement entropy for the outside should decrease; hence, $S_{\mathcal{EL}} < S_{\mathcal{E}}$. Then, inconsistency is encountered because $S_{\mathcal{L}} < 0$, which is impossible.

The point that we mentioned is that after the Page time, no annihilation of antiparticles can occur. If this is the case, then the entanglement entropy between black hole and radiation cannot decrease after the Page time. Therefore, we obtain the relation $S_{\mathcal{EL}} > S_{\mathcal{E}}$. Now we can avoid the inconsistency of the AMPS argument.

Based on the above, we need to change Fig.~\ref{fig:Fig} after the Page time. The infalling antiparticles cannot find their partners. We assume that (1) before the Page time, the annihilations of antiparticles are maximally allowed, (2) Hawking radiation is a creation of separable pairs, and (3) non-local interactions occur. Then, after the Page time, $S_{\mathcal{EL}} > S_{\mathcal{E}}$ should be satisfied. In fact, as we have seen in Fig.~\ref{fig:sim3}, an addition of Hawking particles will increase the entanglement entropy slightly. Thus, the entanglement entropy should monotonically increase, though it never reaches the Page limit. At the same time, due to the accumulation of antiparticles inside the horizon, the total number of states will monotonically increase. The corresponding entropy diagram is summarized in Fig.~\ref{fig:newpage}. Note that this picture is consistent with the entanglement entropy behavior of a moving mirror that mimics the remnant scenario \cite{Chen:2017lum}. Also, a similar form is expected from another point of view \cite{Hotta:2015huj}.

Now we summarize the modified scenario in Fig.~\ref{fig:Fig_mod}. In the first step, we reach the maximum entangled state. Hence, the numbers of degrees of freedom between inside and outside are approximately equal. In this context, let us consider that one more Hawking particle is created. In the second step, the incoming antiparticle cannot interact with the entire system. Therefore, a limitation on the ability of the antiparticle to find its partner particle appears. If this is impossible, then the antiparticle will be retained without further censorship or annihilation. Then in the last step, at the end of evaporation, the total number of degrees of freedom is twice the original one.

Outside is usual radiation, but inside is a mixture of half the original collapsed matter and antiparticles. The total energy of the black hole has now vanished or is at least negligible, but the final mixture should maintain its entanglements with radiation.

\subsection{Interpretations}

Before the Page time, in principle, one can trace out the degrees of freedom of antiparticles inside the black hole, though it requires several strong assumptions. Indeed, this is the only way to maintain the Bekenstein-Hawking entropy formula, $\log n = \mathcal{A}/4$, where $\mathcal{A}$ is the area of the event horizon (if one cannot erase the number of states inside the black hole, then the number of states should increase even though the areal entropy decreases). If this is possible, then the total number of degrees of freedom can be maintained, and the entanglement entropy between the black hole and radiation will monotonically increase according to the Page formula.

After the Page time, this is not possible. Therefore, approximately four possibilities, where this analysis is quite similar to that of \cite{Ong:2016iwi}, can occur:
\begin{itemize}
\item[--] After the Page time, the ingoing antiparticles disappeared via non-unitary processes. This is related to the loss of information \cite{Unruh:1995gn}, at least effectively \cite{Maldacena:2001kr,Sasaki:2014spa}. Also, this picture is related to the Horowitz-Maldacena proposal \cite{Horowitz:2003he}.
\item[--] Non-local interactions may occur between inside and outside the horizon \cite{Page:2013mqa}.
\item[--] After the Page time, Hawking radiation cannot be a creation of separable particle pairs. This is related to the firewall conjecture \cite{Almheiri:2012rt}.
\item[--] After the Page time, antiparticles are stored inside the horizon, and the total number of degrees of freedom increases. This is related to the (broadly defined) remnant scenario \cite{Chen:2014jwq}.
\end{itemize}

By assuming the last possibility (Fig.~\ref{fig:newpage}), as the number of radiation degrees of freedom increases, the number of internal degrees of freedom will increase by the same amount as that of radiation. Our observation based on numerical investigation shows that even though antiparticles are interacting with internal degrees of freedom, the total system is no longer in a random state (Fig.~\ref{fig:sim3}). Hence, as the number of internal degrees of freedom increases, the entanglement entropy between inside and outside cannot reach the Page limit.

Note that for a unitary process, we can define three kinds of information \cite{Alonso-Serrano:2015bcr}:
\begin{eqnarray}
I(A) &=& \log n - S(A|B),\\
I(B) &=& \log m - S(B|A),\\
I(A:B) &=& S(A|B) + S(B|A) - S(A \cup B),
\end{eqnarray}
where for unitary systems, $S(A|B) = S(B|A)$ and $S(A \cup B) = 0$. Hence, $I(A) + I(B) + I(A:B) = \log nm$, where $\log nm$ is the maximum information capacity for given $n$ and $m$. This formula will hold for time-varying $n$ and $m$. We call $I(A)$ and $I(B)$ the information measures for the black hole and the radiation \cite{Lloyd:1988cn}, respectively, and $I(A:B)$ the mutual information between the black hole and the radiation. This gives two important results.
\begin{itemize}
\item[1.] Even though we consider such an idealized scenario to erase states inside the horizon, the Bekenstein-Hawking entropy bound, $\log n = \mathcal{A}/4$, must be broken around the Page time. This is the way toward the remnant picture. However, based on our toy model calculations, we can specify that such a breakdown should happen even at times around the Page time, where the black hole is still big and semi-classical.
\item[2.] Eventually, information will be stored in the form of $I(A)$, $I(B)$, and $I(A:B)$. Note that Hawking particles after the Page time cannot make the state random enough. Therefore, non-trivial $I(B)$ is found after the Page time. This means that one can distinguish information from Hawking radiation after the Page time, which is the same as black hole complementarity \cite{Susskind:1993if}. However, whether an inconsistency via the information duplication argument like black hole complementarity exists is less clear because in this model, we counted inside and outside the black hole consistently. (However, once the entanglement entropy begins to decrease, one can find an inconsistency via an argument similar to one in \cite{Chen:2014jwq}.)
\end{itemize}
By measuring Hawking radiation, one can, in principle, check the amount of information and can distinguish whether this scenario is true or not.

The following is an important and new conclusion: \textit{if we assume unitarity, the semi-classical quantum field theory, and general relativity, then the black hole should violate the Bekenstein-Hawking entropy bound at times around the Page time at the latest}. If the remnant scenario is correct, then the remnant or over-entropy object should appear even in the semi-classical limit; hence, in some sense, the \textit{remnant should, in general, be large}. Then, the infinite production argument of remnants \cite{Giddings:1994qt} can be applied to semi-classical over-entropy objects, which looks more severe.

\section{Discussion}\label{sec:dis}

In this paper, we investigated interacting bipartite systems based on several toy models. Our main question was this: how can we erase states inside the horizon in a unitary way? The only way is to find a separable pair inside the horizon. Even if we use strong assumptions, this is possible only before the Page time; after the Page time, it is impossible, which is a conclusion consistent with the AMPS argument.

Then, what does this mean? By assuming unitarity and Hawking pairs as separable systems, the Bekenstein-Hawking entropy bound should breakdown at times around the Page time, which can correspond to a large black hole. Thus, we were able to obtain a consistent and extended Page diagram for remnants (Fig.~\ref{fig:newpage}), but this caused various obstacles, including the infinite production problem.

On the other hand, what happens if we allow an assumption that Hawking pairs are no longer separable after the Page time? This is related to the firewall conjecture, but we can further emphasize that the role of the firewall should prevent the creation of separable pairs. However, Hawking particles are created not only at the horizon but also far outside the horizon, e.g., $\sim 3 M$ \cite{Davies:1976ei}. Then, if such a creation of separable pairs is to be prevented, the firewall should be located way outside the event horizon. Then, this firewall must be severely naked \cite{Hwang:2012nn}, which can be easily ruled out.

This may indicate that the easiest way to reconcile all the inconsistencies is to lose information and allow non-unitary evolution, at least, effectively. Of course, we need to explain how to realize this, but we postpone that for a future project. In any case, our investigations based on simple toy models give deep insights into the information loss problem and make clearer the fact that the information loss problem is not easy to resolve based on the current state of our knowledge.

\begin{acknowledgments}
The authors would like to thank Pisin Chen, Masahiro Hotta, and Jiunn-Wei Chen for valuable discussions during the 3rd LeCosPA symposium. DY was supported by the Korea Ministry of Education, Science and Technology, Gyeongsangbuk-Do, and Pohang City for Independent Junior Research Groups at the Asia Pacific Center for Theoretical Physics and the National Research Foundation of Korea (Grant No.: 2018R1D1A1B07049126). This work was supported by a Daegu Gyeongbuk Institute of Science and Technology (DGIST) Undergraduate Group Research Project (UGRP) grant. 
\end{acknowledgments}

\end{document}